
%
\baselineskip 14pt
\voffset=1.5cm
\hoffset=1.0cm
\vsize=21.5cm
\hsize=6in

\font\tenpt=cmr10
\font\twelvept=cmr10 scaled 1200
\font\twelvebf=cmb10 scaled 1200
\font\twelvesl=cmsl10 scaled 1200
\def\etal{{\sl et al.} }
\def\ratio#1#2{{{#1}\over{#2}}}
\tolerance=5000 \parindent=0pt\parskip=6pt \baselineskip14pt \tenpt

\def\ref{\par\noindent\hangindent=52pt\hangafter=1}

\def\bff{\twelvebf}
\def\sll{\twelvesl}
\def\blankline{\par\vskip 8pt}

\def\etal{\sll et al.}

\def\lapp{\mathbin{\raise2pt \hbox{$<$} \hskip-9pt \lower4pt
\hbox{$\sim$}}}
\def\gapp{\mathbin{\raise2pt \hbox{$>$} \hskip-9pt \lower4pt
\hbox{$\sim$}}}

\twelvept

\def\rw#1#2{\hbox to size{\!
{\hbox to 4cm{\lft{#1}}}{\hbox to 6cm{\lft{#2}}} \hss }}

\def\el1{\delta (\delta -1)}
\def\ta1{\eta (\eta -1)}

\par\noindent
\centerline{\bff Dynamical and radiative properties of
astrophysical supersonic jets}
\par\vskip.5cm\noindent
\centerline{\bff I.  Cocoon morphologies}
\par\vskip1.cm\noindent
\centerline{S. Massaglia$^{(1)}$, G. Bodo$^{(1)}$, A.
Ferrari$^{(1),(2)}$}
\twelvept
\par\vskip0.6cm\noindent
\par\noindent
$^1${\tensl Osservatorio Astronomico di Torino, Strada dell'Osservatorio
20, I -- 10025, Pino Torinese -- Italy}
\par\noindent
$^2${\tensl Istituto di Fisica Matematica dell'Universit\`a,
Via C. Alberto 10, I -- 10123, Torino -- Italy}
\par\vskip1.cm\noindent
{\bff Abstract.}
We present the results of a numerical analysis of the propagation
and interaction of a supersonic jet
 with the external medium.
We discuss the motion of the head of the jet into the
ambient in different physical conditions, carrying out calculations
with different Mach numbers and  density ratios of the jet to the exteriors.
 Performing the calculation in
a reference frame in motion with the jet head, we can follow in detail its
 long  term dynamics. This numerical scheme allows us also to
study the morphology of the cocoon for different physical parameters.
We find that the propagation velocity
of the jet head into the ambient medium strongly influences
the morphology of the cocoon, and this result
 can be relevant in connection to the origin and structure
of lobes in extragalactic radiosources.
 \blankline\noindent
 \blankline\noindent
{\bff 1. Introduction}
 \blankline
Since the  pioneering work by Norman et al.\  (1982), many numerical studies
have been devoted to the analysis of the propagation of a  supersonic jet shot
into an ambient medium.  The first studies (Norman et al.\ 1982, 1983, 1984;
 Wilson \& Scheuer 1983) showed the main features of the interaction between
jet and environment. We can describe the basic picture which emerged from
those results in
the following way: the deceleration of the jet flow at its head is
accomplished through the formation of a strong shock (Mach disk) which
thermalizes the jet bulk kinetic energy; the overpressured shocked jet material
forms a backflow along the sides of the jet and inflates a cocoon whose
size increases decreasing the density ratio between jet and ambient material;
finally, a second shock (bow shock) is driven into the external medium. This
basic picture bore also a suggestive resemblance with the structures seen in
radio maps of extragalactic jets and brought to the association of the compact
hot
spots with the working surfaces where the jet dissipates its kinetic energy and
of the radio lobes with the cocoons formed by the jet waste material. The
following studies introduced many different ingredient to the basic model in
order to make it more similar to the real astrophysical situation (for a recent
review see e.g. Burns, Norman \& Clarke 1991). These ingredients include
 variability  of the injection properties of the jet (Clarke \& Burns 1991),
 variation of the physical parameters of the ambient medium
along the jet propagation path (Norman, Burns \& Sulkanen 1988)
 and nonadiabaticity of the flow, relevant to the case of stellar jets
(Blondin, Fryxell \& K\"onigl 1990). There have been also attempts
to study the fully 3-D case (Norman, Stone \& Clarke 1991,
Hardee \& Clarke 1992, Hardee, Clarke \& Howell 1995) and to introduce MHD
effects (Clarke, Norman \& Burns 1986; Lind et al.\ 1989).
More recently, numerical simulations of relativistic jets have been carried out
by Mart\'{\i}, M\"uller \& Ib\'a\~{n}ez (1994) and Duncan \& Hughes (1994),
for low Mach number jets, and by Mart\'{\i} et al. (1995) for high Mach number
jets.

In spite of these strong efforts many aspects of this problem are still not
well
understood. This is due to the complexity of the jet-cocoon structure: in fact,
the
cocoon excites perturbations to the jet flow, which in turn can be amplified by
the Kelvin--Helmholtz mechanism and induce a strong activity of the jet's head
that
affects the cocoon structure. Thus a complex feedback loop mechanism
establishes
between jet and cocoon which make the dynamics of the interaction very complex.
In addition, when trying a more direct comparison
of the results with observations, one must remember that
what is observed is an outcome of the distribution of energetic particles
and magnetic field
and not the bulk of the flowing plasma,
and therefore direct comparisons could be misleading.
In this paper and in a companion one
(Massaglia et al.\  1995, henceforth Paper II) we try to elucidate some of
these aspects.
Here we focus on the dynamics of the interaction and we describe some
 properties of the cocoon structure which can be
relevant for the observational properties of extragalactic radio--sources
and have been overlooked in previous studies. We have been able to examine
these properties because of the
wide exploration of the parameter space especially
towards high Mach numbers,
 typically higher
than those discussed in the present literature
(see however Loken et al. 1992 for
 one simulation with Mach number in the range considered
here),
and because our approach
allowed us to follow the jet propagation up to very long times and to keep
all parts of the cocoon in the computational domain, whereas the usual
approach is limited to follow the jet only for one crossing time of the grid
and to lose the back part of the cocoon.
In the companion Paper II,
 we focus instead on the radiative properties
following the distribution of a passive magnetic field and
 of relativistic particles subject to
synchrotron losses and to adiabatic expansion.

The outline of the paper is the following. In the next Sec. II we
discuss the physical problem; the numerical scheme is examined in Sec. III;
the results of the simulations of the jet's head
are reported in Sec. IV; the application of the results to
astrophysical jets in radio sources are discussed  in Sec. V.
\blankline\noindent
\blankline\noindent
{\bff 2. The Physical Problem}
\blankline
We study the dynamics of a supersonic, cylindrical, axisymmetric
 jet continuously injected in a medium initially
at rest. We
solve numerically the full set of adiabatic, inviscid fluid equations
for mass, momentum, and energy conservation,
$$
{{\partial \rho} \over {\partial t}} + \nabla \cdot (\rho {\vec{v}}) = 0
\,,
$$
$$
\rho {{\partial {\vec{v} }} \over {\partial t}} + ({\vec{v}} \cdot
\nabla){\vec{v} } = -\nabla p \,,
$$
$$
{{\partial p} \over {\partial t}} + ({\vec{v}} \cdot \nabla)p -
\Gamma{p \over \rho} \left[
{{\partial \rho} \over {\partial t}} + ({\vec{v}} \cdot \nabla) \rho
\right] = 0
$$
where the fluid variables $p$, $\rho$ and
$\vec{v}$
are, as customary, the pressure, density, and velocity, respectively;
$\Gamma$ is the ratio of the specific heats.

In order to follow
 the jet particles in the external environment,
we solve an additional advection equation for a scalar field $f$:
$$
{{\partial f} \over {\partial t}} + ({\vec{v} } \cdot \nabla) f = 0
\,.
$$
The initial spatial distribution for this tracer is designed to
demarcate the jet alone; thus, we set $f$ initially equal to one inside
the jet, and to zero outside; in the following evolution, $f$ is set to
one also for the newly injected jet fluid.
 By this means, we can distinguish between
the matter which is initially part of the jet or is afterwards injected
in the jet, and that which is part
of the external medium.

Similar calculations found in the literature
(see, e.g., Norman et al. 1982, 1984; Lind et al. 1989) are done
 injecting  the jet from the left boundary in a medium at rest.
In this way i) one is limited to follow the evolution
of the jet's head only up to the time when it reaches the right boundary of the
computational grid, and ii) it is not possible to
determine the geometrical structure resulting from
the interaction of the jet with the ambient medium
since the back part of this structure is lost out of the left boundary
of the domain. Since our main goal is to follow the long term evolution
of this structure, we have
overcome this difficulty by  carrying out  the actual computations  in a
reference frame in which the jet's head is nearly at rest and well
inside the computational domain (see Fig. 1). Therefore,
 in the initial configuration,  the external medium moves at a uniform velocity
$-V_{\rm h}$, where
$$
V_{\rm h} = {v_{\rm j} \over 1+\sqrt{\nu}} \;,
$$
where $v_{\rm j}$ is the jet velocity in the `laboratory frame' and $\nu$ is
the
ratio of the external to the jet density,
and $V_{\rm h}$ is an approximated advance velocity of the jet's head
(see below \S 4.2).
This moving frame is adopted in the computations, but afterwards,
we will discuss the results obtained, translating them
back in the reference frame where the external medium is at rest, i.e.
in the `laboratory frame'.

Thus, the jet initially  occupies a
a cylinder of length $L$ (see Fig. 1), in pressure equilibrium
with the external medium and
the initial flow structure  has  the following form:
$$
v_z(r)=
 \left\{
\eqalign{
    {{v_z(r=0)} \over
   {\cosh[({r})^m]}}- V_{\rm h}  \phantom{-V_{\rm h}}
  ,\quad & z\le L \,\,, \cr
                     -V_{\rm h}
 \phantom{    {{v_z(r=0)} \over
    {\cosh[({r})^m]}}-V_{\rm h} }
 ,\quad & z > L \,\,. \cr}
\right.
\;,
$$
where $m$ is
a `steepness' parameter for the shear layer separating the jet from the
external medium.
The choice of separating the jet's interior from the ambient medium with a
smooth
transition, instead of a sharp discontinuity,  avoids numerical instabilities
that can develop at the interface between the
jet's proper and the exteriors, expecially at high Mach numbers.

The density radial dependence has the form:
$$
{\rho(r) \over \rho(r=0)} = \nu - \ratio{ \nu-1 } {\cosh[({wr})^n]} \,.
$$
We have carried out a series of calculations setting $w=0.75$, $m=8$
and $n=2m$; this implies a narrower and smoother radial
 extension of the `density' jet with respect to the `velocity' jet.
 The reason for this choice is to obtain a smooth
 radial profile of the momentum density $\rho v_z$.

\blankline
\blankline
{\bff 3. The Numerical Scheme}
\blankline\noindent
{\it 3.1.) Scaling}
\blankline\noindent
An important step in order to determine the relevant control parameters
for this problem, and therefore the extension of the parameter space to be
explored,
is to non-dimensionalize the system of equations.
In this case, we have chosen to measure all lengths in units of the jet radius
$a$,
 and time in units of
the sound crossing time $t_{\rm cr}= a/v_{\rm sound}$ ($v_{\rm sound}$ is the
initial isothermal sound speed on the jet's axis: $r=0$, $t=0$). Also  density
and pressure are expressed in units of their values at $r=0$ and
$t=0$ while the velocities are expressed in terms of $v_{\rm sound}$.
With this choice of non-dimensionality, the control parameters are then reduced
to the jet (internal) Mach number
$$
M = {v_z(r=0,t=0) \over \sqrt{\Gamma} v_{\rm sound}} \;,
$$
and the density ratio $\nu$.
\blankline\noindent
{\it 3.2.) Integration domain and boundary conditions}
\blankline\noindent
Integration is performed in cylindrical geometry and
the domain of integration ($0 \leq z \leq D$,  $0
\leq r \leq R$) is covered by a grid of $ 750 \times 250$ grid points.
The axis of the beam is taken coincident with the bottom boundary
of the domain ($r=0$), where symmetric (for $p$, $\rho$ and $v_z$) or
antisymmetric (for $v_r$) boundary conditions are assumed. At the top
boundary ($r=R$) and right
boundary ($z=D$) we choose free outflow conditions, imposing
for every variable $Q$ null gradient ($dQ/d(r,z)=0$).
These free conditions
do not completely avoid back-reflection phenomena from the outer
boundaries. In order to limit this effect, the boundaries should be placed
as far as possible from the region of the jet where the most interesting
evolutionary effects presumably take place;
for this purpose we employ
a nonuniform grid both in the longitudinal $(z)$ and the radial $(r)$
directions
(Fig. 1).
In the radial direction the grid is uniform over the first 50 points and then
the
mesh size is increased assuming $\Delta r_{j+1} = 1.015 \Delta r_{\rm j}$. In
this
way the
jet spans over 20 uniform meshes, while the external boundary is
shifted to about $r \simeq 66$.
as for the $z$--direction, we assume constant grid in the
central part of the domain, i.e. in a sub-domain of length 40,
 between 150 and
600 grid points; conversely, in the remaining part we consider an
expanded  grid increasing the mesh
distance according to the scaling law $\Delta z_{j \pm 1} = 1.015 \Delta z_{\rm
j}$,
where the minus sign applies in the first 150 grid points and the plus sign
above 600 grid points.  This non-uniform grid
has high resolution in the region where the jet's head
is maintained by the co-moving reference frame; at the same time
it has the advantages to
 place the boundaries as far as
possible and to allow to study the backflow and the cocoon structure for a
longer time, before boundary reflection effects set in.
As a comparison, Loken et al. (1992) adopted a $1200 \times 350$ non uniform
grid
in the radial direction, while Blondin \& Cioffi (1992) used a $600 \times 300$
uniform grid. The design of our grid allows us to achieve a better resolution
in
the central region of the domain and a coarser grid
in the peripheral parts with respect to Loken et al. (1992); however, with our
approach,  we manage to keep most of the cocoon in this central region.
Conversely, Blondin \& Cioffi (1992) obtain a higher resolution but with a
smaller size of the domain.

The numerical scheme adopted is of PPM (Piecewise Parabolic Method) type and
is particularly well suited for studying highly supersonic flows with strong
shocks (Woodward \& Colella 1984, see also Bodo et al. 1994, 1995).

\blankline
\blankline
{\bff 4. Dynamical evolution}
\blankline\noindent
{\it 4.1.) General features}
\blankline\noindent
The general features of the evolution of a jet's head propagating into an
external medium have been widely
described in the literature since the presentation of the first
simulations by Norman {\etal} (1982, 1984) and, in order to put our
results in the full context,   we summarize
them here briefly. The early phases are
essentially related to the unfolding of the initial discontinuity between
jet and external material into i) a reverse shock propagating in the jet
against
the flow, ii) a contact  discontinuity, separating the jet material from the
external medium, and  iii) a shock propagating in the external material. We
can therefore distinguish five different regions  in the evolved
structure: 1) the jet proper; 2) the shocked jet material  still flowing
in the forward direction; 3) the shocked jet material reflected backwards
at the contact discontinuity and flowing back at the  jet side; 4) the
shocked external material;  5) the unshocked external material. The
shocked jet and external material forms an expanding overpressured
region which is called cocoon. The high pressure cocoon squeezes the jet
and drives into it shock waves, which reflecting on the axis assume the
characteristic biconical shape seen in the simulation results.
These shocks modify the structure
of the jet head and affect its propagation into the ambient medium.
This complex interaction is the object of our investigation and will
be discussed in detail in the following sections. The results of the
interaction depends on the parameter  $M$ and $\nu$, which define the jet,
and  therefore the
morphology of the cocoon will be determined by the choice of these parameters.

The strength of the shock waves, driven by the cocoon into the jet, and the jet
squeezing depend on how much the cocoon is overpressured.  The cocoon pressure
is in turn determined by the  kinetic energy flow, thermalized at
the jet shock, and  by the cocoon expansion. We therefore expect a higher
pressure for jet of higher Mach number and lower density: in this case,
in fact, the energy flow into the cocoon is higher. This is confirmed by
our numerical results that show a stronger dependence on the Mach number and
a weaker dependence on the density ratio, in agreement with the simplified
analytical model by Begelman \& Cioffi (1989). A stronger interaction between
biconical shocks and jet head is therefore expected for high Mach number jets
and it is in fact in this parameter range that our results show different
behaviours related to this interaction. The jet thrust can be modulated
by the biconical shocks impinging on its head and this can produce a periodic
increase in the advance velocity of the head, leading to a strong
change in the cocoon morphology.
In order to study in detail these processes, we have carried out
several simulation runs exploring the parameter plane $(\nu,M)$
for underdense, hypersonic jets.
Fig. 2 shows the effective extent to which we have explored the
$(\nu,M)$-parameter space.  Different symbols (bullets and circles)
refer to the different behaviors mentioned above and discussed in more
detail below.
In subsection 4.2 we will examine
in detail how the advance velocity is modified by the jet perturbations,
while in subsection 4.3 we will examine the effects on the cocoon morphology.

 \blankline\noindent
{\it 4.2.)  Velocity of the jet's head}
\blankline\noindent
The advance velocity of the jet's head can be estimated on a first
aproximation by balancing the ram-pressure exerted by the jet front,
$\rho_{\rm j} (v_{\rm j}-V_{\rm h})^2 $, with the analogous force exerted
by the external medium $\rho_{\rm ext} V_{\rm h}^2 $ (see  e.g. Norman
et al. 1982, 1984). The equilibrium condition between these two forces
gives the head velocity as
$$
 V_{\rm h} \simeq {v_{\rm j} \over
1+\sqrt{\nu }} \;,
\eqno(1)
 $$
Since this value will serve in the following for comparison, it will
be indicated by $V_{\rm h}$, while the actual head velocity obtained from the
numerical
results will be indicated by $v_{\rm h}$.
The numerical results
obtained by many authors give in general a velocity lower than  $V_{\rm h}$.
  Lind et al.\ (1989) give an interpretation for this lower
velocity, noting that the areas on which the two ram pressures exerts can
be in general different and this amounts to changing Eq. 1 as follows:
$$
V_{\rm h} \simeq {v_{\rm j} \over 1+\sqrt{\nu / \epsilon }} \;,
$$
where $\epsilon=A_{\rm j}/A_{\rm h}$ takes into
account the expansion (or contraction) of the jet at the head.
Since, in general, the jet head tends to expand,
its advance velocity will be correspondingly lower than that estimated
 by Eq. 1.

However, as discussed above, the interaction of the biconical shocks with
the jet head can lead in some cases to a contraction of the  head area
and correspondingly to an advance velocity larger than that predicted by
Eq. 1. In order to discriminate between these two behaviours, we have
plotted in Fig. 3, the distance $z_{\rm h}$, covered by the jet's head,
as a function of the  normalized time $\tau$ defined below,
 for jets with different values of the parameters $M$ and $\nu$.

 A typical difficulty one has to face
in comparing results obtained with different parameters $\nu$ and $M$ is
that the evolution proceeds at a different pace for every set of
parameters, i.e. comparable configurations can occur at different epochs.
Following Cioffi \& Blondin (1992), we treat this problem introducing
  a `normalization' time, $t_{\rm norm}$, defined
as the time employed to cover the unit distance (i.e. the
jet radius) moving at the velocity $V_{\rm h}$ (Eq. 1).
 Configurations at the same
$\tau =t/t_{\rm norm}$ are, within reasonable approximation, in a similar
evolutionary stage and can then be compared. We will therefore make use of
this normalized time $\tau$ in every plot representing the temporal
evolution of some quantity.
In our units defined in Section 3.1, we have $t_{\rm
norm}^{-1} = M/(1+\sqrt{\nu})$ and $\tau=t M / (1 + \sqrt{\nu})$.

The use of the normalized time $\tau$ in Fig. 3 makes very easy to distinguish
between jet's head propagating at velocities above or below the value given
by Eq. 1. In fact, this value corresponds to the line  $z_{\rm h} = \tau$.
In the figure we have a group of curves lying above this line, indicating
larger velocities and a group of curves lying below, indicating lower
velocities. The higher velocities are found for high Mach number jets with
low values of $\nu$, while in all the other cases the velocities are lower and
tend to decrease with time as already discussed by various authors.

We can gain a better understanding of the differences between the two classes
looking at the behavior of the head  velocity $v_{\rm h}$
 as a function of
time. The three
panels in Fig. 4 show the results obtained for the cases $M=100$, $\nu=10$;
$M =3$, $\nu = 10$ and
$M=100$, $\nu=100$. The first case belongs to the high (head) velocity class,
while the other two cases belong to the low velocity class.
We can immediately notice a great difference in behaviour between the first
case
and the other two:
in panel a) we see that the velocity oscillates almost periodically, staying
always above the value $V_h$ showed for comparison as a dashed line; in panels
b) and c) the velocity decreases systematically
showing irregular oscillations of low amplitude
and is well below the value $V_h$ for case b).

 We can follow in more detail two events of increase of the head velocity
through a sequence of contour plots of the distribution of the intensity
of the longitudinal momentum flux  $\rho v_z^2$ in the $r,z$ plane
(Fig. 5). The time
of each frame of the sequence can be traced on the velocity plot (Fig. 4).
Frames a) and b) show  a strong (nonlinear) perturbation, excited by the
backflow interacting with
the jet's wall; it  travels towards the jet axis, reaching it and causing a
steepening of ram pressure at the axis that accelerates the
head's advance.
In fact the first maximum of Fig. 4a occurs after
the time employed by the perturbation to
cover one jet radius. As time elapses, this perturbation is reflected at the
axis, reaches the edge of the jet (frames c, d) and is reflected back again
towards the axis and reaches it in the
head's region (frames e, f), with the consequent increase of $v_{\rm h}$ that
leads
to the second maximum, that in fact corresponds to the time employed by the
perturbation to cover twice the jet radius. The same reasoning applies for the
following maxima.

 Fig. 6 represents a sequence of contour plots
  of the distribution of the intensity
of the longitudinal momentum flux, as in Fig. 5 above, but for a
 low Mach number jet, e.g. $M=3$ and $\nu=10$. We can clearly
 see the formation of the biconical shocks and their interaction
 with the jet head.   The relevant difference is that the strength
  of the ram pressure perturbations results much lower than in the
  previous case. This can be more clearly  seen looking at the plots
  of the same quantity as a function of the longitudinal
  coordinate $z$, on the jet axis,
  presented in Fig. 8: the periodic
  increases in momentum flux are present also in this case (panel b) but they
  have a much lower amplitude. As discussed in Section 4.1, in the low Mach
  number regime we expect a much lower cocoon pressure and therefore a much
  weaker perturbation induced in the jet by the cocoon. Fig. 9 shows the
  temporal behaviour of the average cocoon pressure and we can note the
  low values for this case (panel b) compared with the other two high
  Mach number cases.
 Following the evolution of the jet's head in the sequence
  of images in Fig. 6, we notice that it tends to expand further weakening
  the effect of the biconical shocks,  with a
consequent slower progress in the ambient medium.

 The two classes discussed above are shown with different symbols in the Fig. 2
representing the explored parameter plane. Bullets (region 1)
indicate those
cases for which the head velocity $v_{\rm h}$ presents large quasi periodic
oscillations and
is in the average larger than $V_{\rm h}$; empty circles (region 2), instead,
indicate the cases in which $v_{\rm h} < V_{\rm h}$, with irregular low
amplitude oscillations. In this distribution we can note that the high velocity
class
is bounded at low Mach numbers and at very low jet densities compared to the
external medium (high $\nu$).
The same process of generation of biconical shocks, which in turn interact
with the jet head is at work also in the low velocity region 2 of the parameter
plane but it is not able to impart sufficient momentum to the head
and increase in a sensible manner its velocity. The reasons why
this does not happen are however different for the low Mach number jets and
for the low density jets.

We investigate at this point why the low density cases ($\nu > 30$) present
characteristics similar to
a low velocity behavior. Looking at Figs. 7c and 8c for the case $M=100$,
$\nu=100$, in fact, we see large
values of the cocoon pressure and consequently of the momentum flux
perturbations, but, on the contrary, the head velocity does not show
corresponding large variations. The answer can be found looking again at the
sequence of images of the momentum flux distributions for this case, presented
in Fig. 7. Comparing these images with the corresponding images for the case
$M = 100$, $\nu = 10$ (Fig. 5), we see that in the lower density case the
biconical shocks form a larger angle with the jet axis: the jet compression is
therefore immediately followed by a strong expansion and the increased thrust
acts on the  head  for a very short time only. The momentum imparted to
the jet's head is therefore low and the same is true for the increase in
velocity.
In the higher density case, on the contrary, the biconical shocks forms a small
angle with the jet axis, the enhanced thrust acting on the head can last longer
and the resulting increase in velocity is greater. The critical parameter
appears
therefore to be the inclination of the biconical shocks and this in turn
appears
to depend mainly on the density
ratio $\nu$: for low values of $\nu$ we find small angles between the shocks
and
the jet axis and the angle increases increasing $\nu$. This explains why we
find
the borderline for the high velocity behaviour at $ \nu \simeq 30$
(Fig. 2).

\blankline\noindent
{\it 4.3.) Cocoon morphology}
\blankline\noindent
As discussed before, the particular setup chosen allows to
study how the morphology of the cocoon varies in different regions of the
parameter plane, and to compare then with the observed morphologies of radio
sources. Figs. 10a,b represents two typical morphologies corresponding to the
two regions of different behavior of the head velocity:
  the bimodal behaviour of
$v_{\rm h}$, discussed previously, reflects on this morphology.
We measure the shape of the
cocoon, given by the contour of the
tracer $f > 0$, as shown in Fig. 11:  here $L$ is the longitudinal size of the
cocoon,
$Y_{\rm M}$ the maximum radius, and $X_{\rm M}$ the position of this
maximum radius with respect to
the jet's head. The ratios $Y_{\rm M}/L$ and $X_{\rm M}/L$ characterize the
different morphologies.

In Fig. 12a) we show $Y_{\rm M}/L$ against $\tau$; from this figure we note
that, as
time elapses, the values attained by the ratio $Y_{\rm M}/L \simeq 0.2 - 0.4$,
with a
tendency for high $M$ to produce low values of  $Y_{\rm M}/L$.
Begelman \& Cioffi (1989) (see also Loken et al. 1992
for a more general formulation) have given estimates of the temporal
evolution of the cocoon ratio $Y_{\rm M}/L$, under the assumption
of constant $v_{\rm h} \ll v_{\rm j}$ for a cylindrical cocoon. They find that
the ratio varies with time as $\propto t^{-1/2}$; in Fig. 12a) stars
show this analytical result as a reference. More interesting is  the temporal
behaviour of
the ratio $X_{\rm M}/L$ for the different parameters (Fig. 12b). Here the
bimodal
structure is again apparent: jets in region 1 of the parameter plane  produce
cocoon with elongated morphologies ($X_{\rm M}/L \simeq 0.6$) as in Fig. 10b)
(`spearhead' cocoon),
while jets in region 2  lead to cocoon morphologies that present
the largest radius towards the head   ($X_{\rm M}/L \simeq 0.2$) as in the
example of
Fig. 10a) (`fat' cocoon).

This different behaviour can be interpreted examining the variation of the
shape of
the cocoon during the episodes of sudden acceleration of the jets in region 1.
Two
of these episodes are represented in the two panels of Fig. 13, and we see
that, for each
velocity increase, the jet's head surpasses the position of the old bow shock
creating a
new smaller bow shock yielding a global V shape. For jets in region 2, instead,
this
does not happen and the cocoon tends to expand at its front, giving a
completely different morphology.

Comparing the obtained morphologies to the simulation by Loken et al. (1992),
we
note a qualitative difference in the shape of the cocoon. We recall that we
consider an initial pressure balance between the jet and the ambient, while
Loken et al. (1992) study an overpressured jet. However, we think that the main
reason lies in the different initial setups chosen: in our case it is possible
to follow the fate of the backflow due to the head interaction,
i.e. the cocoon is fully contained in the domain, while in the
case of Loken et al. (1992), and also Blondin \& Cioffi (1992), the backflow is
partially lost through the left boundary since the very beginning.

We finally add a comment about the limits of a
2-D geometry. We know that in 2-D large scale structures tend to be favoured,
 therefore some of the features one finds in 2-D simulations could not
form, or be unstable,  in a complete 3-D treatment. However
 the qualitative
behaviour of the interaction can be captured in a simplified 2-D
geometry. In fact, 3-D simulations by Hardee \& Clarke (1992)
and Hardee, Clarke \& Howell (1995) show that key features, such
as biconical shocks, are present also in a fully 3-D
geometry.
\blankline\noindent
\blankline\noindent
{\bff 5. Astrophysical applications and conclusions}
\blankline\noindent
We have discussed the physical characteristics of the interaction of
supersonic,
underdense, cylindrical fluid jets with a homogeneous, undisturbed medium. In
order to
examine the different cocoon morphologies, as they evolve in time, we have
employed a particular setup that allows to follow this
evolution on long time scales. The application of the results obtained to
extragalactic jets
can be performed relating the morphologies emerging from the numerical
calculations to those observed mainly in the radio band. A basic question that
arises at this point is the following: which physical quantity is most suitable
to represent the observed radio brightness distribution? Authors have usually
considered the particle density as a tracer of the brightness distribution,
especially as far as radio lobes were concerned, drawing conclusions on the
bases of how this quantity `looked like' in a given numerical experiment and
comparing this to the astrophysical situation. Instead,
the brightness distribution is a function of the relativistic particle density
distribution in space and energy and on the magnetic field. In the present
calculations we do not study these quantities leaving
this aspect for a
 forthcoming Paper II. However we can already attempt to
identify some quite general trend looking at the behaviour of the tracer $f$,
that we recall provides snapshots of the spatial distribution of the jet
particles.

 Looking at the tracer distributions in simulation runs with different values
of $M$ and $\nu$, we have noted a typical bimodal behaviour that we have
interpreted on the basis of the temporal, and spatial, evolution of the
longitudinal momentum flux $\rho v_z^2$. Thus, from the
point of view of the cocoon morphology, slow jets and fast jets with high
density ratios behave differently from fast jets with low density ratios,
as it is clearly apparent on Fig. 11b). Can this fact be, at least in a very
broad sense, related to different radio lobe morphologies? In other words,
is the cocoon representative of radio lobes?

To attempt  answering this question, we compare the density distribution
in Figs. 10a,b) to the tracer distribution as given in Figs. 14a,b), for the
same values of $M$ and
$\nu$.
We recall that $f$ suffers from numerical diffusion in its evolution
and, albeit being initially 1 inside and 0 outside the jet, it assumes values
intermediate between these two extremes as turbulent effects develop.
However, as discussed in Bodo et al. (1995), $f$ remains
a good marker of the jet particles. Going back to Figs. 10 and 14, we note
that,
for $M=100$,
the shock that surrounds the cocoon involves  matter of the
external ambient. Is this shocked region site of particle acceleration? If the
answer is positive we can say that the form of the lobe resembles the density
distribution of Fig. 10b), with an elongated structure having the front part
protruding from the lobe. Similar morphologies can be found in the
sample of high liminosity radio sources by
Leahy \& Perley (1991); representative examples can be: 3C42
($(X_{\rm M}/L)_{\rm obs} \sim 0.75$), 3C184.1 ($\sim 0.8$), 3C223 ($\sim
0.7$),
3C441 ($\sim 0.7$), 3C349 ($\sim 0.77$), 3C390.3 ($\sim 0.6$). In this scheme,
the jet would be
characterized by a high value of the Mach number accompanied by a moderate
value
of the density ratio. Moreover, the shock that surrounds the cocoon,
according to simulations, must have
the effect of enhancing the component of the magnetic field along the shock
front, resulting in a high  polarization at the source
edge, with the polarization vector directed
normally to the edge itself. This effect is clearly visible in the polarization
maps of the sources mentioned above.
In case of absence of particle acceleration in the
external shock, and always assuming that the relativistic particled are carried
along
the jet following the tracer $f$, the `lobe' would look like Fig. 14b), i.e.
rather unusual to observations.

In the case of slow jet, we note from Fig. 10a) that the shock forms only in
the
front part of the cocoon, therefore the actual lobe has to have a morphology
similar to that given by the tracer in Fig. 14a). Examples of this second kind
of
morphology can be found in the sample of Leahy \& Perley (1991): 3C296
($(X_{\rm M}/L)_{\rm obs} \sim 0.34$), 3C296  ($\sim 0.33$) and
3C173.1 ($\sim 0.32$)
are good examples of this class, while 3C382  ($\sim 0.4$) and 3C457 ($\sim
0.25$)
 are less so; the
remaining sources of the sample are more irregular and many appear to bear some
characteristics of both kind. We note however that a detailed comparison must
allow for
 possible projection effects: $(Y_{\rm M}/X_{\rm M})_{\rm model} \ll
(Y_{\rm M}/X_{\rm M})_{\rm obs}$.

\blankline\noindent
{\it Acknowledgements. The numerical calculations reported
in the paper have been carried out by the
CRAY YMP computer of Consorzio per il Supercalcolo del Piemonte.}

\blankline\noindent
\blankline\noindent
{\bff References}
\blankline\noindent
\ref
Begelman M. C., Cioffi D. F., 1989, ApJ 345, L21

\ref
Bodo G., Massaglia S., Ferrari A., Trussoni E., 1994,
A\&A 283, 655

\ref
Bodo G., Massaglia S., Rossi P., Rosner R., Malagoli A., Ferrari A., 1995,
A\&A, in press

\ref
Blondin J. M., Fryxell B. A., K\"onigl A., 1990, ApJ 360, 370

\ref
Burns J. O., Norman M. L., Clarke D. A., 1991,
Sci 253, 522

\ref
Cioffi D. F., Blondin J. M., 1992, ApJ 392, 458

\ref
Clarke D. A., Norman M. L., Burns J. O., 1986, ApJ 311, L63

\ref
Clarke D. A., Burns J. O., 1991, ApJ 369, 308

\ref
Hardee P. E., Clarke D. A., 1992, Apj 400, 9

\ref
Hardee P. E., Clarke D. A., Howell D. A., 1995, Apj 441, 644

\ref
Leahy J. P., Perley R. A., 1991, AJ 102, 537.
ApJ 344, 89

\ref
Lind, K. R., Payne, D. G., Meyer, D. L., Blandford, R. D., 1989,
ApJ 344, 89

\ref
Loken C., Burns J. O., Clarke D. A., Norman M. L., 1992, ApJ 392, 54

\ref
Mart\'{\i} J. M$^{\underline{\rm {a}}}$, M\"uller E., Ib\'a\~{n}ez  J.
M$^{\underline{\rm {a}}}$, 1994, A\&A 281, L9

\ref
Mart\'{\i} J. M$^{\underline{\rm {a}}}$, M\"uller E., Font, J. A., Ib\'a\~{n}ez
 J.
M$^{\underline{\rm {a}}}$, 1994, A\&A 281, L9

\ref
Massaglia S., Bodo, G., Rossi P., Ferrari A., 1995,
(Paper II) in preparation

\ref
Norman, M. L., Smarr, L., Winkler, K. H. A., Smith, M. D., 1982,
A\&A 113, 285

\ref
Norman, M.L., Winkler, K.H.A., Smarr, L., 1983, in: Astrophysical Jets,
eds.\ A. Ferrari, A. G. Pacholczyk, Reidel, Dordrecht, p. 227

\ref
Norman, M. L., Winkler, K. H. A., Smarr, L., 1984, in: Physics
of Energy Transport in Extragalactic Radio Sources, NRAO
Worksh. 9, eds.\ A. H. Bridle, J. A. Eilek, p. 150

\ref
Norman, M. L., Burns, J. O., Sulkanen, M. E., 1988, Nat 335, 146

\ref
Norman M. L., Stone J. M., Clarke D. A.,  1991,
AIAA, Aerospace Sciences Meeting, 29th, Reno, NV, Jan. 7-10, p. 12

\ref
Wilson M. J., Scheuer P. A. G., 1983, MNRAS 205, 449

\ref
Woodward, P. R., Colella, P., 1984, J. Comp. Phys. 54, 174
\blankline\noindent
\blankline\noindent
\par\vfill\eject
\par\noindent
{\bff Figure captions}
\blankline\noindent
Figure 1.
Computational domain defined by $0 < z <
D$ and $0 < r < R$, where $D = 139$ and $R = 66$. In the left panel,
we show in black the initial jet area (defined by $0 < z < L$ and
$0 < r < a$), symmetric with respect to the $r=0$ axis.
The right panel displays the non-uniform
computational grid at scale $1:25$.
\blankline\noindent
Figure 2.
{Coverage of the parameter space: symbols indicate the cases in the
$(\nu,M)$-plane which we have actually computed. The different choice of
symbols
(bullets and circles)
indicates regions with different behavior in the jet head advance
velocity and in the cocoon morphology.
              }
\blankline\noindent
Figure 3.
 Plot of distance covered by the jet's head $z_{\rm h}$ as a function
of the normalized time $\tau$. The values of the parameters are reported
in the legenda.

\blankline\noindent
Figure 4.
{ Plot of head's velocity $v_{\rm h}$ vs $\tau$ for $M=100$ and $\nu=10$
(solid line), the dashed line gives $V_{\rm h}$ as a comparison (panel a);
The same as panel a but for $M=3$ and $\nu=10$   (panel b);
The same as panel a but for $M=100$ and $\nu=100$.   (panel c).
              }
\blankline\noindent
Figure 5.
{  Sequence of contour plots of the distribution
of the intensity of the longitudinal momentum flux $\rho v_z^2$
in the $r - z$ plane, for the case $M = 100$, $\nu = 10$.
The time of each frame is marked on the velocity
plot in Fig. 4a.
              }
\blankline\noindent
Figure 6.
{  Sequence of contour plots of the distribution
of the intensity of the longitudinal momentum flux $\rho v_z^2$
in the $r - z$ plane, for the case $M = 3$, $\nu = 10$.
The time of each frame is marked on the velocity
plot in Fig. 4b.
              }
\blankline\noindent
Figure 7.
{  Sequence of contour plots of the distribution
of the intensity of the longitudinal momentum flux $\rho v_z^2$
in the $r - z$ plane, for the case $M = 100$, $\nu = 100$.
The time of each frame is marked on the velocity
plot in Fig. 4b.
              }
\blankline\noindent
Figure 8.
{ Plot of the on axis
longitudinal momentum flux profile $\rho v_z^2$,
scaled with respect to the on axis initial value, for $M=100$ and
$\nu=10$ (panels a, b) and for  $M=3$ and $\nu=10$ (panels c, d), at the time
corresponding to the first and second maxima of $v_{\rm h}$ for $M=100$.
              }
\blankline\noindent
Figure 9.
{Plot of the temporal behavior of the average cocoon pressure
vs $\tau$ for $M=100$ and $\nu=10$  (panel a);
The same as panel a but for $M=3$ and $\nu=10$   (panel b);
The same as panel a but for $M=100$ and $\nu=100$ (panel c).  Dashed lines
represent the power-law best fit $P_c\propto \tau ^{-\eta}$ to the
average cocoon pressure, to be compared to the analytical estimate $\eta=1$
by Begelman \& Cioffi (1989).           }
\blankline\noindent
Figure 10.
{Gray scale image of the density for $M=10$ and
$\nu=100$ (panel a, `fat' cocoon), and $M=100$ and $\nu=10$ (panel b,
`spearhead' cocoon). This in an example of
the two characteristic structures assumed by the cocoon.
              }
\blankline\noindent
Figure 11.
{Measuring a cocoon.
              }
\blankline\noindent
Figure 12.
{Temporal evolution of $Y_{\rm M}/L$ (panel a) and of $X_{\rm M}/L$
(panel b). The different sets of parameters are reported in legenda.
              }
\blankline\noindent
Figure 13.
{Gray scale images of pressure distribution for the case $M =100$,
$\nu =10$ at two different times showing the effects of the head velocity
variations on the cocoon morphology.}
\blankline\noindent
Figure 14.
{The same as in Fig. 10, but with the tracer.
              }
\bye